\documentclass[journal,onecolumn]{IEEEtran}
\UseRawInputEncoding
\usepackage{bm}
\usepackage{graphicx}
\usepackage{subfigure}
\usepackage{float}
\usepackage{booktabs}
\usepackage{caption}
\usepackage{epstopdf}
\usepackage{epsfig}
\usepackage{algpseudocode}
\usepackage{multirow}
\usepackage{booktabs}
\usepackage[normalem]{ulem}
\usepackage{lineno}
\useunder{\uline}{\ul}{}
\usepackage{threeparttable}
\usepackage{cite}
\usepackage{enumerate}
\usepackage{amsmath}
\usepackage{color}
\usepackage{amssymb}
\usepackage{algorithm}

\begin{document}
\title{Spectral Variability Augmented Sparse Unmixing of Hyperspectral Images}

\author{Ge~Zhang,~
        Shaohui~Mei,~
        Mingyang Ma,~
        Yan~Feng,~
        and~Qian~Du~
\thanks{This work is supported by National Natural Science Foundation of China (62171381) and the Fundamental Research Funds for the Central Universities. Corresponding author: Shaohui Mei (meish@nwpu.edu.cn).}
\thanks{Ge Zhang, Shaohui Mei, Mingyang Ma, and Yan Feng are with the School of Electronics and Information, Northwestern Polytechnical University, Xi'an 710129, China.}
\thanks{Qian Du is with the Department of Electrical and Computer Engineering, Mississippi State University, Starkville, MS 39762, USA.}
}


\maketitle

\begin{abstract}
Spectral unmixing (SU) expresses the mixed pixels existed in hyperspectral images as the product of endmember and abundance, which has been widely used in hyperspectral imagery analysis.
However, the influence of light, acquisition conditions and the inherent properties of materials, results in that the identified endmembers can vary spectrally within a given image (construed as spectral variability).
To address this issue, recent methods usually use a priori obtained spectral library to represent multiple characteristic spectra of the same object, but few of them extracted the spectral variability  explicitly.
In this paper, a spectral variability augmented sparse unmixing model (SVASU) is proposed, in which the spectral variability is extracted for the first time. The variable spectra are divided into two parts of intrinsic spectrum and spectral variability for spectral reconstruction, and modeled synchronously in the SU model adding the regular terms restricting the sparsity of abundance and the generalization of the variability coefficient.
It is noted that the spectral variability library and the intrinsic spectral library are all constructed from the In-situ observed image.
Experimental results over both synthetic and real-world data sets demonstrate that the augmented decomposition by spectral variability significantly improves the unmixing performance than the decomposition only by spectral library, as well as compared to state-of-the-art algorithms.
\end{abstract}

\IEEEpeerreviewmaketitle
\section{Introduction}
\IEEEPARstart{W}{ith} the rapid development of remote sensing technology, hyperspectral image (HSI) collecting abundant spectral information with hundreds of contiguous bands, has been widely used in different kinds of applications, such as geoinformation science, space research, material mapping, and geological exploration \cite{Ghamisi2017Advances}. However, the phenomenon of mixed pixels is widely existed due to insufficient spatial resolution of imaging system, changeable atmospheric environment, and complex distribution of ground objects. The decomposition of a mixed pixel into spectral signatures (endmembers) with corresponding proportions (abundances) is known as hyperspectral unmixing \cite{Bioucas2012Hyperspectral}.

Linear mixing model (LMM) has been widely used for spectral mixture analysis as its computational tractability and flexibility and it supposes that the spectra collected by the imaging spectrometer are a linear combination of endmembers weighted by their corresponding abundance fractions \cite{Heinz2001FCLS}.
However, nonlinearity and spectral variability affect the performance of spectral unmixing \cite{Borsoi2021Review}.
Techniques based on geometry, statistics, and nonnegative matrix factorization extract endmembers directly from the HSI.
In general, an endmember extraction step is applied, and then, abundance value for each pixel is estimated. There are many algorithms for endmember extraction, such as N-FINDR \cite{Winter1999NFINDR}, pixel purity index (PPI) \cite{boardman1995mapping}, and vertex component analysis (VCA) \cite{nascimento2005vertex}. Nevertheless, these algorithms require pure pixel assumption, and it is not always satisfied due to the spatial resolution. Others of them extract virtual endmembers without physical meaning in a given HSI \cite{Li2015Minimum}.

Recently, due to the wide availability of spectral libraries, like the U.S. Geological Survey (USGS) digital spectral library, a semi-supervised spectral unmixing approach, called sparse regression-based method, has been shown to circumvent the drawbacks introduced by such virtual endmembers and the unavailability of pure pixels \cite{Iordache2011Sparse,Mei2021Sparse}. It assumes that the mixed pixels can be expressed in the form of linear combinations of a small number of pure spectral signatures from a (potentially very large) spectral library known in advance.
Another significant advantage is that it does not need to extract endmembers from the hyperspectral data or estimate the number of the endmembers. Generally, the number of endmembers in the scene is small compared with the number of endmembers in the spectral library, which means that only a small number of endmembers contribute to the mixed pixel. Therefore, the abundance vector of the mixed pixel estimated by sparse unmixing is expected to be sparse. These new perspectives introduced by sparse unmixing fostered advanced developments in the field \cite{Xu2021Pruning,Ma2021Dictionary,Mei2021BlockSparse}.

In the past few years, several algorithms have been developed to enforce the
sparsity on the solution of SU \cite{Qi2020Multiview,Zhang2020Scaling}.
The sparse unmixing algorithm via variable splitting and augmented Lagrangian (SUnSAL) \cite{Iordache2011Sparse} adopts the $L_1$ regularizer on the abundance matrix to measure the sparsity of the abundance vector in
each pixel. The collaborative SUnSAL (CLSUnSAL) \cite{Iordache2014Collaborative} introduces the $L_{2,1}$ regularizer to constrain the pixels in HSIs to share the same active set of endmembers and ensure that all the abundance vectors exhibit global row sparsity. Based on the alternating direction method of multipliers (ADMM) \cite{Dias2010admm}, the corresponding sparse regression problems are all convex and can be settled efficiently.
Moreover, the spatial information acts a pivotal part in sparse unmixing as the corresponding problem is usually made easier on a local scale \cite{Ince2021Spatial}.
Total variation (TV) \cite{Iordache2012Total,Qin2021Total,Feng2020Total} regularization exploits the spatial information via a first-order pixel neighborhood system to significantly improve the unmixing results, but it may yield oversmoothness and blurred boundaries.
Zhang \emph{et al.} \cite{Zhang2018S2} proposed a spectral¨Cspatial weighted sparse unmixing (S$^2$WSU) framework constrained simultaneously from the spectral and spatial domains to further imposing sparsity on the solution.
To obtain spatial-contextual information, a fast multiscale spatial regularization unmixing algorithm (MUA) was proposed by Borsoi \emph{et al.} \cite{Borsoi2019MUA}. In this algorithm, superpixel-based segmentation is performed before unmixing to form a coarse domain, and the unmixing result for the coarse domain forms a multiscale spatial regularization for unmixing of the original domain.
Another solution is to focus on the local spectral similarity of the HSIs rather than the simple proximity of the location.
The local spectral similarity preserving (LSSP) constraint \cite{Li2019local} was proposed to preserve spectral similarity in a local region given that adjacent pixels share not only the same endmembers with high probability but also approximated fractional abundances.
Meanwhile, the low-rank constraint of the abundance matrix has been increasingly adopted for sparse unmixing, providing a new perspective for spatial correlation \cite{Sun2020Rank}.
Giampouras \emph{et al.} \cite{Giampouras2016ADSpLRU} simultaneously imposed single sparsity and low rankness on abundance matrices, taking into account both sparsity and spatial correlation information in HSIs.
The joint-sparse-blocks and low-rank unmixing (JSpBLRU) algorithm \cite{Huang2019JSpBLRU} further imposed the joint-sparsity-blocks structure and low rankness on abundance matrices for pixels in a sliding window.
Recently, the sparse unmixing method named superpixel-based reweighted low-rank and total variation (SUSRLR-TV) \cite{Li2021Superpixel} was proposed by using superpixel segmentation to extract the homogeneous regions, and imposing a low-rank constraint to promote the correlation of each superpixel s abundance matrix.

However, the spectral signatures of the materials contained in hyperspectral images can be significantly affected by spectral variability, which results from different imaging conditions including atmospheric effects, illumination, topographic changes, and the intrinsic variation of the spectral signatures of the materials (i.e., due to physicochemical differences), especially in a hyperspectral image with a higher spatial resolution \cite{Drumetz2020Extended,Drumetz2020Convex,Theiler2019Causes}. Therefore, the LMM hardly makes an accurate unmixing in reality, due to such ubiquitous error that passively transfer the unpredicted errors into LMM \cite{borsoi2020multiscale}.
The existing methods could be basically divided into two groups: 1) dictionary or bundle-based approaches \cite{Uezato2019Bundles,Drumetz2019Bundles}, which try to model endmembers by a certain number of instances of each material, 2) model-based approaches, which describes the endmember variability
by a specific statistical distribution \cite{Halimi2015Unsupervised,Liu2021Bayesian,zhou2020probability} or by incorporating the variability in the mixing model based on physically motivated concepts \cite{thouvenin2015hyperspectral,Drumetz2016Extended,Hong2019Augmented}. This work is developed following the first approach.

The key idea in dictionary or bundle-based approaches consists in extracting several instances of each endmember in order to build a dictionary, which is then used for spectral unmixing, offering more than one representative spectral signature per material \cite{Zhang2019Assessing,Drumetz2016Recent}.
Representing each endmember class with large spectral libraries of spectra as a \emph{priori} does not require the assumption of any particular spectral distribution within a single endmember class.
Methods for this category select the appropriate spectra, or bands, from different spectral sets according to pre-defined criteria for weakening the effect of endmember variab  ility, which can be roughly divided into four groups: multiple endmember spectral mixture analysis (MESMA), sparse unmixing, spectral transformations, and machine learning.

The basic principle behind MESMA \cite{roberts1998mapping} is to iteratively search for the combination of endmember signatures in the library that, among all possibilities, enables the closest reconstruction of each observed pixel under the LMM. The MESMA algorithm and its variants formulate spectral unmixing as a computationally demanding optimization problem and achieve good quality \cite{xu2016image}.
Chen \emph{et al.} \cite{Chen2016Multiple} proposed a sparse multiple-endmember spectral mixture model (SMESMM) by using a block sparse algorithm to obtain an initial block sparse solution and resolving the mixed pixel using the selected land cover materials.
Zhang \emph{et al.} \cite{zhang2016multiscale} incorporated spatial information in MESMA by using segmentation algorithms to divide the image into different homogeneous objects, which are then unmixed individually by using a library that is also constructed from object based spectra.
Borsoi \emph{et al.} \cite{Borsoi2020Generative} leveraged the power of deep generative models to learn the statistical distribution of the endmembers available in the existing libraries, and then drew new samples to augment the spectral libraries, improving the overall quality of the unmixing process.

Compared to MESMA, sparse unmixing formulations use mathematical relaxations that are computationally easier to solve. A typical approach \cite{berman2017comparison} was to modify the LMM for unmixing mineral spectra in mining applications by including an additional term representing the mixture of the background spectrum of the endmembers. This background spectrum was defined as the low-frequency part of the spectral signatures and estimated a priori from the library as a parametric function of smooth splines. The performance of an $L_{1}$ norm-based sparse unmixing framework under this model was reported to be similar to MESMA, albeit at a much smaller computational cost.

In the spectral transformations group, instead of the original reflectance data, transformed spectral information with less spectral variability are used as input to the SMA, like derivative spectra and wavelet transformed spectra \cite{Singh2012transformation}. However, spectral transformations are empirically oriented techniques and require a significant degree of expert knowledge about the underlying application \cite{shao2019derivative}.
Other types of methods are more machine learning-oriented, which allow to learn the function linking the endmembers available in the dictionary (or in a supervised fashion using an a priori available dictionary). The training can be performed for instance by simulating mixed pixels with endmembers from the dictionary in different proportions \cite{Mianji2011SVM}, or models the latent function between spectra and abundances in a training set and predicts abundances from a given pixel spectrum under Gaussian process framework \cite{Uezato2016guass}. Machine learning algorithms provide more flexible ways but these methods have the drawbacks of not explicitly modeling the variability, and at a large computational complexity.

Traditional spectral unmixing (SU) algorithms neglect the spectral variability of the endmembers, which propagates significant modeling errors throughout the whole unmixing process and compromises the quality of the results.
The recent library-based SU methods address spectral variability by using libraries of spectra that had to be acquired a priori (e.g., through in situ measurements) without introducing explicit spectral variability term, which limited the applicability of these approaches.
Therefore, a spectral variability augmented sparse unmixing model (SVASU) is proposed, which provides a new way to construct spectral variability explicitly from the spectral library for sparse unmixing.
Finally, extensive experiments over both synthetic and real-world datasets are conducted to validate the effectiveness and robustness of the proposed SVASU method for hyperspectral sparse unmixing.

The remainder of this paper is organized as follows. Section II introduces the basic idea of sparse unmixing.
Section III presents the proposed SVASU model for SU of hyperspectral images.
Section IV reports and discusses experimental results on both synthetic and real datasets.
Finally, Section V draws a conclusion.

\section{Sparse Unmixing}
\begin{figure*}[!t]
\begin{center}
\includegraphics*[width=18cm]{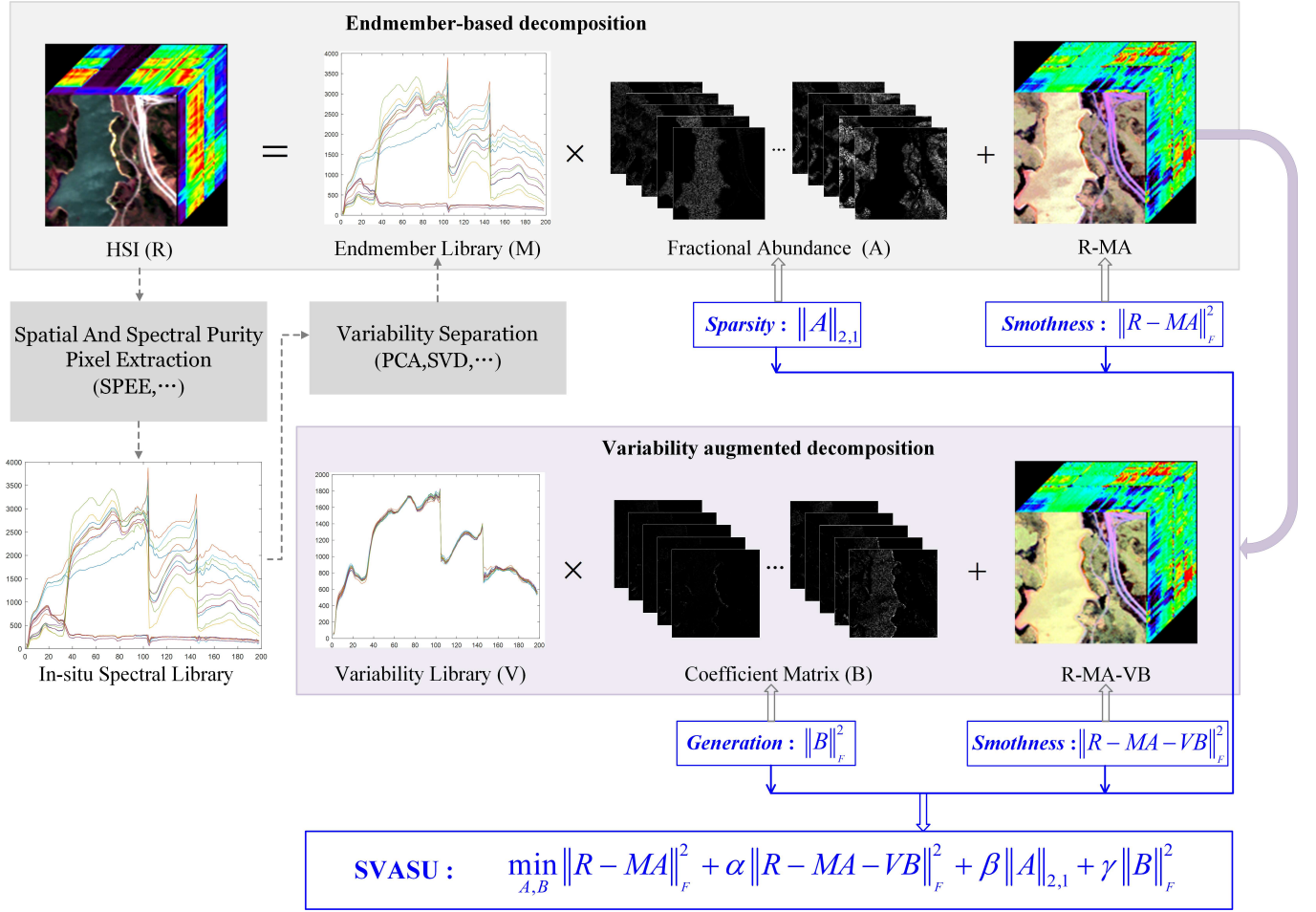}
\end{center}
\caption{Basic flowchart of sparse unmixing through SVASU.}
\label{flowchart}
\end{figure*}
\subsection{Linear Mixture Model}
The Linear Mixture Model (LMM), which is well-known for its
simplicity and explicit physical meaning, has been widely utilized
to model the relationship between mixtures and endmembers. In the
LMM, the photons reflected from different ground objects within one pixel are assumed not to interfere with each other. As a
result, the observed hyperspectral image matrix $\mathbf{R} \in \mathcal {R}^{b
\times n}$ is assumed to be the product of endmember matrix
$\mathbf{E} \in \mathcal {R}^{b \times p}$ and its corresponding
abundance matrix $\mathbf{A} \in \mathcal {R}^{p \times n}$ plus a
noise matrix $\mathbf{N} \in \mathcal {R}^{b \times n}$. Their relationship can be
formulated as:
\begin{equation} \label{eqn_LMM}
\mathbf{R}=\mathbf{EA}+\mathbf{N},
\end{equation}
in which $b$ is the number of bands, $p$ is the number of
endmembers, and $n$ is the number of pixels in the image. Here, the
$i$-th column vectors of $\mathbf{E}$ and $\mathbf{A}$, denoted by
$\mathbf{e}_i$ and $\mathbf{a}_i$, correspond to the $i$-th
endmember and the abundance of the $i$-th spectral pixel
$\mathbf{r}_i$, respectively.

In order to make the LMM physically meaningful,
the abundance sum-to-one constraint (ASC) and the abundance nonnegative constraint (ANC) must be imposed, which can be expressed as:
\begin{equation}
\label{eqn_ASC} \sum_{i=1}^p{a_{ij}}=1,
\end{equation}
\begin{equation}
\label{eqn_ANC} a_{ij} \geq 0 \qquad i=1,2,\ldots,p, j=1,2,\ldots,n.
\end{equation}
Generally, the ASC can be embedded into this model by adding an
additional pseudo band to the data matrix and the endmember matrix
\cite{Mei2010Mixture}. Therefore, in the following analysis, only
the ANC is explicitly considered.

\subsection{Sparse Unmixing}
Due to the wide availability of spectral libraries, sparse unmixing models use mathematical relaxations that are computationally easier to solve.
Let $\mathbf{M} \in \mathbb{R}^{b \times m}$ be a large spectral library containing $m$ spectral signatures, and $\mathbf{A} \in \mathbb {R}^{m \times n}$ denotes the abundance maps corresponding to library $\mathbf{M}$ for the observed data $\mathbf{R}$.
As the number of endmembers involved in a mixed pixel is usually very small when compared with the size of the spectral library, the abundance matrix $\mathbf{A}$ contains many zero values, which means $\mathbf{A}$ is sparse.
Thus, the unmixing problem can be formulated as an $l_2-l_0$ optimization problem,
\begin{equation}\label{eqn_l2l0}
\mathop {\min }\limits_\mathbf{A} \frac{1} 2 \| \mathbf{R}-\mathbf{MA} \|_F^2 + \lambda \| \mathbf{A}\|_0 \ \ \  \mathrm{s.t.}\ \mathbf{A} \geq  0,
\end{equation}
where $\|\cdot \|_F$ is the Frobenius norm, $\|\cdot \|_0$ is the $l_0$-norm which denotes the number of nonzero components of a vector, and $\lambda$ is a regularization parameter used to adjust the weight of the sparsity in this model.

Problem (\ref{eqn_l2l0}) is nonconvex and difficult to be solved. However, the $l_0$-norm constrained sparsity can be relaxed using $l_1$ norm, by which the sparse unmixing algorithm via variable splitting and augmented Lagrangian (SUnSAL) \cite{Iordache2011Sparse} using the $l_2-l_1$ norm to solve the unmixing problem as follows,
\begin{equation}\label{eqn_l2l1}
\mathop {\min }\limits_\mathbf{A} \frac{1} 2 \| \mathbf{R}-\mathbf{MA} \|_F^2 + \lambda \| \mathbf{A}\|_{1,1} \ \ \  \mathrm{s.t.}\ \mathbf{A} \geq  0,
\end{equation}
where $\|\mathbf{A}\|_{1,1} = \sum_{j=1}^n \|\mathbf{a}_j\|$ with $\mathbf{a}_j$ being the $j$-th column of $\mathbf{A}$.
However, the Problem (\ref{eqn_l2l1}) is heavily influenced by the high correlation of spectral libraries due to the underdetermined nature. To admit a sufficiently sparse solution for sparse unmixing will guarantee a more accurate abundance estimation. In \cite{Iordache2014Collaborative}, the Collaborative Sparse Unmixing via variable Splitting and Augmented Lagrangian (CLSUnSAL) algorithm imposes the joint sparsity with an $l_{2,1}$ mixed norm among the endmembers simultaneously for all of the pixels, whose objective function can be defined as follows,
\begin{equation}\label{eqn_l2l21}
\mathop {\min }\limits_\mathbf{A} \frac{1} 2 \| \mathbf{R}-\mathbf{MA} \|_F^2 + \lambda \sum_{i=1}^m\| \mathbf{a}^i\|_2 \ \ \  \mathrm{s.t.}\ \mathbf{A} \geq  0,
\end{equation}
where $\mathbf{a}^i$ denotes the $i$-th line of $\mathbf{A}$ and $\sum_{i=1}^m\| \mathbf{a}^i\|_2$ is the so-called $l_{2,1}$ mixed norm.

\section{Sparse Unmixing Via SVASU}
A considerable number of such sparse unmixing methods has been developed using libraries of spectra that originally had to be acquired a priori (e.g., through in situ measurements), which used to limit the applicability of these approaches.
An important recent development concerns methods that can extract spectral libraries directly from the observed images or generate them using physics-based mathematical models of material spectra. This supports the widespread applicability of library-based sparse unmixing techniques in situations where spectral libraries are not available or cannot be built.

However, the spectral signature variability affects the performance of sparse unmixing task since that the inaccurate endmember representation undoubtedly results in the propagation of errors in abundance estimates derived using SMA.
Thus, to address spectral variability problem, the spectral variability library is introduced to assist the spectral libraries for better signature approximation in sparse unmixing.
Specifically, the introduction of spectral variability library makes a spectrum signature divided into two independent representation parts, which allows the inherent spectral variability into materials to be represented separately and explicitly.

\subsection{Extraction of In-situ Spectral Library by SPEE}

In a hyperspectral image, pixels from a homogeneous ground object often present in adjacent areas, which result in lots of pure spatial neighborhoods in the image. If all these pure spatial neighborhoods can be detected, their representative spectral signatures can be selected as endmember candidates. Different endmember candidates can be considered as coming from homogeneous ground objects provided that the spatial neighborhoods they represent cover pixels in common \cite{Mei2010Spatial}. Based on this spatial refinement scheme, the remaining spatially independent endmember candidates become more representative since they represent larger areas. The number of endmember candidates to constrcut the in-situ spectral library will also be significantly reduced.
Note that the in-situ spectral library $\mathbf{X}$ is extracted from the observed image $\mathbf{R}$ (refer to \cite{Mei2010Spatial}), containing multiple spectra of each possible material (a total of $p$ materials), written as
\begin{equation}\label{eqn_SPEE}
\begin{aligned}
&\mathbf{X}=[\mathbf{X}^{(1)},\mathbf{X}^{(2)},\cdot\cdot\cdot,\mathbf{X}^{(p)}]\\
&\mathbf{X}^{(i)}=[\mathbf{e}^{(i)}_1,\mathbf{e}^{(i)}_2,\cdot\cdot\cdot,\mathbf{e}^{(i)}_m],\ i=1,2,\cdot\cdot\cdot,p
\end{aligned}
\end{equation}
in which $\mathbf{X}^{(i)}$ represents the library subset corresponding to the $i$-th possible material and is composed of $m$ spectral signatures of the current material.

\subsection{Segmentation of Endmember Library and Spectral Variability Library}

Through the previous section, we have obtained the image spectral library including the spatially independent and spectrally pure pixels in the homogenous region according to the pixel purity. Principal Component Analysis (PCA) is a classic method widely used in signal feature extraction, data dimensionality reduction and compression. It retains the main information of the original image data by retaining the first $k$ principal components corresponding to the larger feature value, and the remaining principal components correspond to noise, spectral variability, and/or outliers.

In this section, PCA is performed on its covariance matrix after standardizing the pixel data $X$ of the in-situ spectral library as
\begin{equation}\label{eqn_PCA1}
\begin{aligned}
\bar{\mathbf{X}} &= \frac{1}{n}\sum_{i=1}^{n}\mathbf{X}_i,\ \ \ \ \ \ \ \ \ \ \
\hat{\mathbf{X}}= \mathbf{X}-\bar{\mathbf{X}},\\
\mathbf{C}&=\frac{1}{n}\hat{\mathbf{X}}\hat{\mathbf{X}}^T,\ \ \ \ \ \ \ \ \ \ \ \ [\mathbf{W},\mathbf{D}]= PCA(\mathbf{C}),\\
\mathbf{W}&=[\mathbf{w}_1,\mathbf{w}_2,\cdot\cdot\cdot,\mathbf{w}_n],\ \
\mathbf{D}=diag(\lambda_1,\lambda_2,\cdot\cdot\cdot,\lambda_n), \ \
\end{aligned}
\end{equation}
where $\lambda_1>\lambda_2>\cdot\cdot\cdot>\lambda_n$ are the eigenvalues of its covariance matrix $\mathbf{C}$, and $\mathbf{W}$ is composed of eigenvectors of the corresponding eigenvalue matrix $\mathbf{D}$.
Among all the principal components, the first $k$ correspond to the unique feature types of various pure pixels in the homogeneous area, while the other principal components correspond to the local spectral variability of the pixels in the homogeneous area.

The determination of $k$ can be achieved by comparing the size of the characteristic value. Set a threshold $\zeta$, if the cumulative eigenvalues corresponding to the first $k$ pivotal elements account for the percentage of the total eigenvalues to reach the threshold:

\begin{equation}\label{eqn_lambda}
\begin{aligned}
{\raise0.7ex\hbox{${\sum\limits_{i = 1}^k {{\lambda _i}} }$} \!\mathord{\left/
 {\vphantom {{\sum\limits_{i = 1}^k {{\lambda _i}} } {\sum\limits_{i = 1}^n {{\lambda _i}} }}}\right.\kern-\nulldelimiterspace}
\!\lower0.7ex\hbox{${\sum\limits_{i = 1}^n {{\lambda _i}} }$}} \ge \zeta .
\end{aligned}
\end{equation}

Therefore, the first $k$ principal components can be calculated by the first $k$ eigenvectors,
\begin{equation}\label{eqn_PCA2}
\begin{aligned}
\mathbf{F}_1 &= \mathbf{W}_k^T \hat{\mathbf{X}},\\
\mathbf{W}_k&=[\mathbf{w}_1,\mathbf{w}_2,\cdot\cdot\cdot,\mathbf{w}_k], k\leq n ,
\end{aligned}
\end{equation}
Then, the reconstructed spectral library by dominant characteristic components is termed as the endmember library,
\begin{equation}\label{eqn_PCA3}
\begin{aligned}
Library_{\rm{endmember}} & = \mathbf{W}\mathbf{F}_1+\bar{\mathbf{X}} = \mathbf{W}_k\mathbf{W}_k^T \mathbf{X}+\bar{\mathbf{X}}.\\
\end{aligned}
\end{equation}
Similarly, the next $n-k$ principal components are constructed as the spectral variability library as
\begin{equation}\label{eqn_PCA4}
\begin{aligned}
\mathbf{F}_2 = \mathbf{W}_{n-k}^T \hat{\mathbf{X}}, \ \ \
\mathbf{W}_{n-k}=[\mathbf{w}_{k+1},\mathbf{w}_{k+2},\cdot\cdot\cdot,\mathbf{w}_n],\\
Library_{\rm{variability}}  = \mathbf{W}\mathbf{F}_2 + \bar{\mathbf{X}} = \mathbf{W}_{n-k}\mathbf{W}_{n-k}^T \hat{\mathbf{X}}+\bar{\mathbf{X}}.\\
\end{aligned}
\end{equation}

\subsection{Spectral Variability Augmented Sparse Unmixing (SVASU)}
Fig. \ref{flowchart} displays the flowchart of proposed SVASU where the well-known Jasper data set is taken as an example. The proposed SVASU adopts a two-order decomposition structure, in which the first order is to decompose the hyperspectral image data into the product of endmember library and abundance fractions roughly, the second order is to further represent the reconstruction error of pixels from first order as the influence of spectral variability library.

Let $\mathbf{V} \in \mathbb{R}^{b \times l}$ be the spectral variability library containing $l$ spectral variability signatures, and $\mathbf{B} \in \mathbb{R}^{l \times n}$ denotes the coefficients corresponding to library $\mathbf{V}$.
Based on the objective function in Eq. (\ref{eqn_l2l21}), an extra spectral variability augmented data fitting term is introduced in the proposed SVASU model, along with the coefficient generalization regularized term, which is expressed as follows,

\begin{equation}\label{eqn_SVASU}
\begin{aligned}
\Theta=\mathop {\min }\limits_{\mathbf{A},\mathbf{B}} {\left\| {\mathbf{R} - \mathbf{MA}} \right\|_F^2} + \alpha {\left\| {\mathbf{R} - \mathbf{MA} - \mathbf{VB}} \right\|_F^2} \\+ \beta {\left\| \mathbf{A} \right\|_{2,1}} + \gamma {\left\| \mathbf{B} \right\|_F^2}
\ \ \ \ \ \ \ \ \  \mathrm{s.t.}\ \mathbf{A} \geq  0,
\end{aligned}
\end{equation}
where $\alpha$ balances the contribution of data fitting from both spectral spectra and spectral variability, $\beta$ and $\gamma$ controls the sparsity of abundance and the generalization of spectral variability coefficient.

Following \cite{nie2010efficient}, we relax the term $\|\mathbf{A} \|_{2,1}$ by $\rm{Tr}(\mathbf{A}^T\mathbf{D}\mathbf{A})$. Thus the function in (\ref{eqn_SVASU}) can be written as
\begin{equation}
\label{eqn_SVASU1}
\begin{aligned}
\Theta=\mathop {\min }\limits_{\mathbf{A},\mathbf{B}} {\left\| {\mathbf{R} - \mathbf{MA}} \right\|_F^2} + \alpha {\left\| {\mathbf{R} - \mathbf{MA} - \mathbf{VB}} \right\|_F^2} \\
+ \beta Tr(\mathbf{A}^T\mathbf{D}\mathbf{A}) + \gamma {\left\| \mathbf{B} \right\|_F^2}
\ \  \mathrm{s.t.}\ \mathbf{A} \geq  0,
\end{aligned}
\end{equation}
where $\mathbf{D} \in \mathbb{R}^{m \times m}$ is a diagonal matrix with the $i$-th diagonal element formulated as follows, in which $\epsilon > 0$ is a stabilization parameter,
\begin{eqnarray} \label{eqn_D}
{\mathbf{D}}_{ii} = \frac{1}{2\sqrt{\mathbf{a}_i^T \mathbf{a}_i+\epsilon}}.
\end{eqnarray}

In order to integrate non-negative constraint conditions into the objective function $\bf{\Theta}$, we introduce Lagrangian multipliers $\delta \in \mathbb{R}_+^{m \times n}$ 
to restrict $\mathbf{A}\geq 0 $.
Therefore, the function (\ref{eqn_SVASU1}) is equivalent to the following function:
\begin{equation}
\label{eqn_SVASU2}
\begin{aligned}
\bf{\Theta}=\mathop {\min }\limits_{\mathbf{A},\mathbf{B}} {\left\| {\mathbf{R} - \mathbf{MA}} \right\|_\emph{F}^2} + \alpha {\left\| {\mathbf{R} - \mathbf{MA} - \mathbf{VB}} \right\|_\emph{F}^2} \\+ \beta {Tr(\mathbf{A}^T\mathbf{D}\mathbf{A})} + \gamma {\left\| \mathbf{B} \right\|_\emph{F}^2} - Tr(\delta \mathbf{A}^T).
\end{aligned}
\end{equation}
When $\mathbf{D}$ is fixed, the original problem can be turned into a general convex optimization problem.
To update all matrixes of the objective function, we use an alternating projected gradient method. By setting the partial derivatives of with respect to $\mathbf{A}$ and $\mathbf{B}$ respectively, we obtain the following function:
\begin{equation}
\label{eqn_deriv}
\left\{
\begin{aligned}
\frac{{\partial \bf{\Theta} }}{{\partial \mathbf{A}}} & = 2((1 + \alpha ){\mathbf{M}^T}\mathbf{MA} - (1 + \alpha ){\mathbf{M}^T}\mathbf{R}\\
& \ \ \ \ \ \ \ \ + \alpha {\mathbf{M}^T}\mathbf{VB} + \beta \mathbf{D}\mathbf{A})- \delta\\
\frac{{\partial \bf{\Theta} }}{{\partial \mathbf{B}}} & = 2(\alpha {\mathbf{V}^T}\mathbf{MA} - \alpha {\mathbf{V}^T}\mathbf{R} + \alpha {\mathbf{V}^T}\mathbf{VB} + \gamma \mathbf{B})
\end{aligned}
\right.
\end{equation}

We take advantage of KKT conditions in this mathematical optimization problem.
Therefore, the following condition should be satisfied:
\begin{equation}
\label{eqn_deriv2}
\left\{
\begin{aligned}
&2((1 + \alpha ){\mathbf{M}^T}\mathbf{MA} - (1 + \alpha ){\mathbf{M}^T}\mathbf{R} \\
& \ \ \ \ \ \ \ \ \ \  + \alpha {\mathbf{M}^T}\mathbf{VB} + \beta (\mathbf{D} + {\mathbf{D}^T})\mathbf{A})_{ik} \cdot \mathbf{A}_{ik} = 0\\
&2(\alpha {\mathbf{V}^T}\mathbf{MA} - \alpha {\mathbf{V}^T}\mathbf{R} + \alpha {\mathbf{V}^T}\mathbf{VB} + \gamma \mathbf{B})_{ij}\cdot \mathbf{B}_{ij} = 0
\end{aligned}
\right.
\end{equation}
Then, we design the following update rules. The convergence of two
update rules will be analyzed in the APPENDIX.
\begin{equation}
\label{eqn_updateA}
\left\{
\begin{aligned}
&\mathbf{A}_{ik}' =\mathbf{\tilde A}_{ik} (\frac{{({\mathbf{M}^T}\mathbf{R} + \alpha \mathbf{M}^T\mathbf{R}- \alpha \mathbf{M}^T\mathbf{VB} )_{ik}}}{{((1 + \alpha ){\mathbf{M}^T}\mathbf{M}\mathbf{\tilde A} + \beta \mathbf{D} \mathbf{\tilde A})_{ik}}})^ \frac{1}{4} \\
&\mathbf{B}_{ij}' = \mathbf{\tilde B}_{ij}  (\frac{{(\alpha {\mathbf{V}^T}\mathbf{R}- \alpha {\mathbf{V}^T}\mathbf{MA})_{ij}}}{{( \alpha {\mathbf{V}^T}\mathbf{V}\mathbf{\tilde B} + \gamma \mathbf{\tilde B})_{ij}}})^ \frac{1}{4} \end{aligned}
\right.
\end{equation}

Therefore, an iterative procedure is adopted.
In each iteration, $\mathbf{A}$ is calculated with the current
$\mathbf{D}$ and $\mathbf{B}$, and then $\mathbf{B}$ and $\mathbf{D}$ are updated based on the current calculated $\mathbf{A}$.
The iterative procedure is repeated until the algorithm converges.
The pseudo code of SVASU is illustrated in Algorithm \ref{alg:SVASU}.

\begin{algorithm}[htb]
\caption{SVASU.}
\label{alg:SVASU}
\begin{algorithmic}[1]
\Require
{Hyperspectral data matrix $\mathbf{R} \in \mathcal {R}^{b \times n}$;
Observed image spectral library $\mathbf{M} \in \mathbb{R}^{b \times m}$;
Spectral variability library $\mathbf{V} \in \mathbb{R}^{b \times l}$.}
\Ensure
Abundance fraction $\mathbf{A}$ and variability coefficient $\mathbf{B}$.\\
\textbf{Initialize} $\mathbf{A} \in \mathbb{R}^{m \times n}$ and $\mathbf{B} \in \mathbb{R}^{l \times n}$ randomly; $t=0$;\\
\textbf{Repeat}:
\State Update
$\mathbf{A}_{t+1} \leftarrow  \mathbf{A}_t \circ (\frac{{{\mathbf{M}^T}\mathbf{R} + \alpha \mathbf{M}^T\mathbf{R}- \alpha \mathbf{M}^T\mathbf{VB}}}{{(1 + \alpha ){\mathbf{M}^T}\mathbf{M}\mathbf{A}_t + \beta \mathbf{D} \mathbf{A}_t}})^ \frac{1}{4}$;
\State Update $\mathbf{B}_{t+1} \leftarrow \mathbf{B}_t \circ (\frac{{\alpha {\mathbf{V}^T}\mathbf{R}- \alpha {\mathbf{V}^T}\mathbf{MA}_{t+1}}}{{ \alpha {\mathbf{V}^T}\mathbf{V}\mathbf{B}_t + \gamma {B}_t}})^ \frac{1}{4}$;
\State Update $\mathbf{D}$;
\State $t=t+1$;\\
\textbf{Until} Convergence;\\
\textbf{Return} $\mathbf{A}$ and $\mathbf{B}$;
\end{algorithmic}
\end{algorithm}

\section{Experiments}
In this section, a series of experiments on both the simulated data set and two real-world data sets (Jasper and Cuprite data sets) is designed to demonstrate the effectiveness of SVASU algorithm for HSIs.
Several well-known sparse unmixing algorithms are adopted for comparison, including SUnSAL \cite{Iordache2011Sparse}, CLSUnSAL \cite{Iordache2014Collaborative}, ADSpLRU \cite{Giampouras2016ADSpLRU}, JSpBLRU \cite{Huang2019JSpBLRU}, MUA$_{\rm{SLIC}}$ \cite{Borsoi2019MUA}.
For all these methods, the coefficient matrices corresponding to each spectral library are all randomly initialized and the values of parameters in each algorithm refer to the reported optimal values.

\subsection{Evaluation Metrics}
To quantitatively evaluate the performance of our proposed SVASU algorithm, the two types of pixel reconstruction error are employed as in the experiments.
Firstly, the root mean square error (RMSE) is used to measure the distance between the true signal and its estimated value, like that of image pixel $\rm{\mathbf{r}}$ and its reconstructed version $\rm{\hat{\mathbf{r}}}$ as $\rm{RMSE_R}$, and true abundance $\rm{\mathbf{a}}$ and its estimated version $\rm{\hat{\mathbf{a}}}$ as $\rm{RMSE_A}$, which are defined as,
\begin{equation}
\begin{aligned}
\label{ReconEr1}
\rm {RMSE_R} = \sqrt {\frac{1}{n}\sum\limits_{i = 1}^n {{{({\mathbf{r}_i} - {{\hat{\mathbf{r}}}_i})}^2}} }.
\end{aligned}
\end{equation}
\begin{equation}
\begin{aligned}
\label{ReconEr2}
\rm {RMSE_A} = \sqrt {\frac{1}{n}\sum\limits_{i = 1}^n {{{({\mathbf{a}_i} - {{\hat{\mathbf{a}}}_i})}^2}} }.
\end{aligned}
\end{equation}
Furthermore, the signal-to-reconstruction-error (SRE) is also adopted to measure the quality of the reconstruction of signals by different algorithms, which can demonstrate more information regarding the power of the error in relation to the power of the signal.
Generally, the higher the SRE (dB) value, the better the performance of the algorithm.
Similarly, we definite $\rm{SRE_R}$ and $\rm{SRE_A}$ as follows,
\begin{equation}
\begin{aligned}
\label{SRE1} {\rm {SRE_R}}= 10\rm{log}_{10} \frac { E[\|\mathbf{r}\|_2^2]} {E[\|{\mathbf{r}} - {{\hat {\mathbf{r}}}}\|_2^2]},
\end{aligned}
\end{equation}
\begin{equation}
\begin{aligned}
\label{SRE2}
{\rm {SRE_A}}= 10\rm{log}_{10} \frac { E[\|\mathbf{a}\|_2^2]} {E[\|{\mathbf{a}} - {{\hat {\mathbf{a}}}}\|_2^2]}.
\end{aligned}
\end{equation}


\begin{figure}[t!]
\begin{center}
\subfigure[Endmember Library]{\includegraphics[width=4cm]{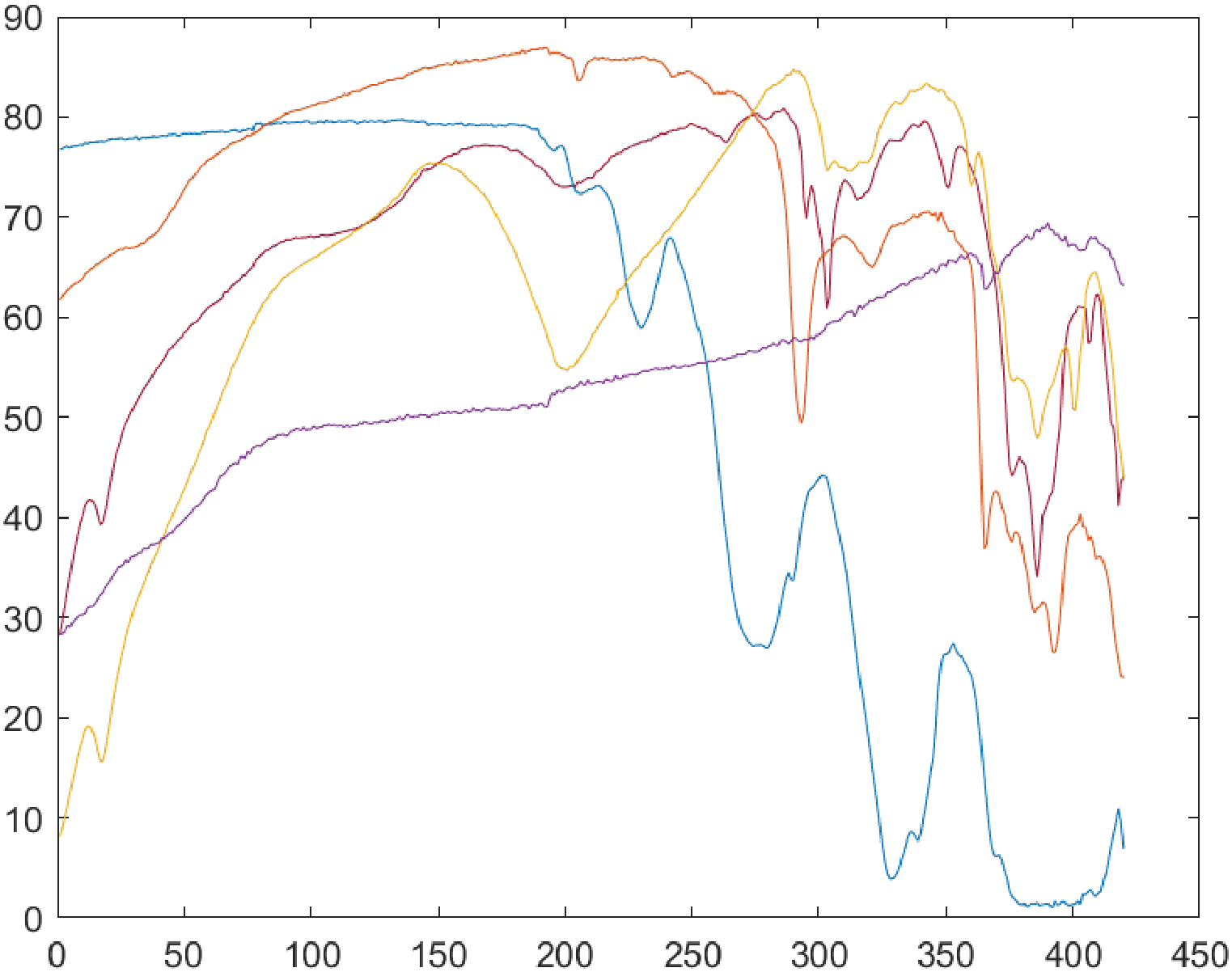}}
\subfigure[Spectral Variability Library]{\includegraphics[width=4cm]{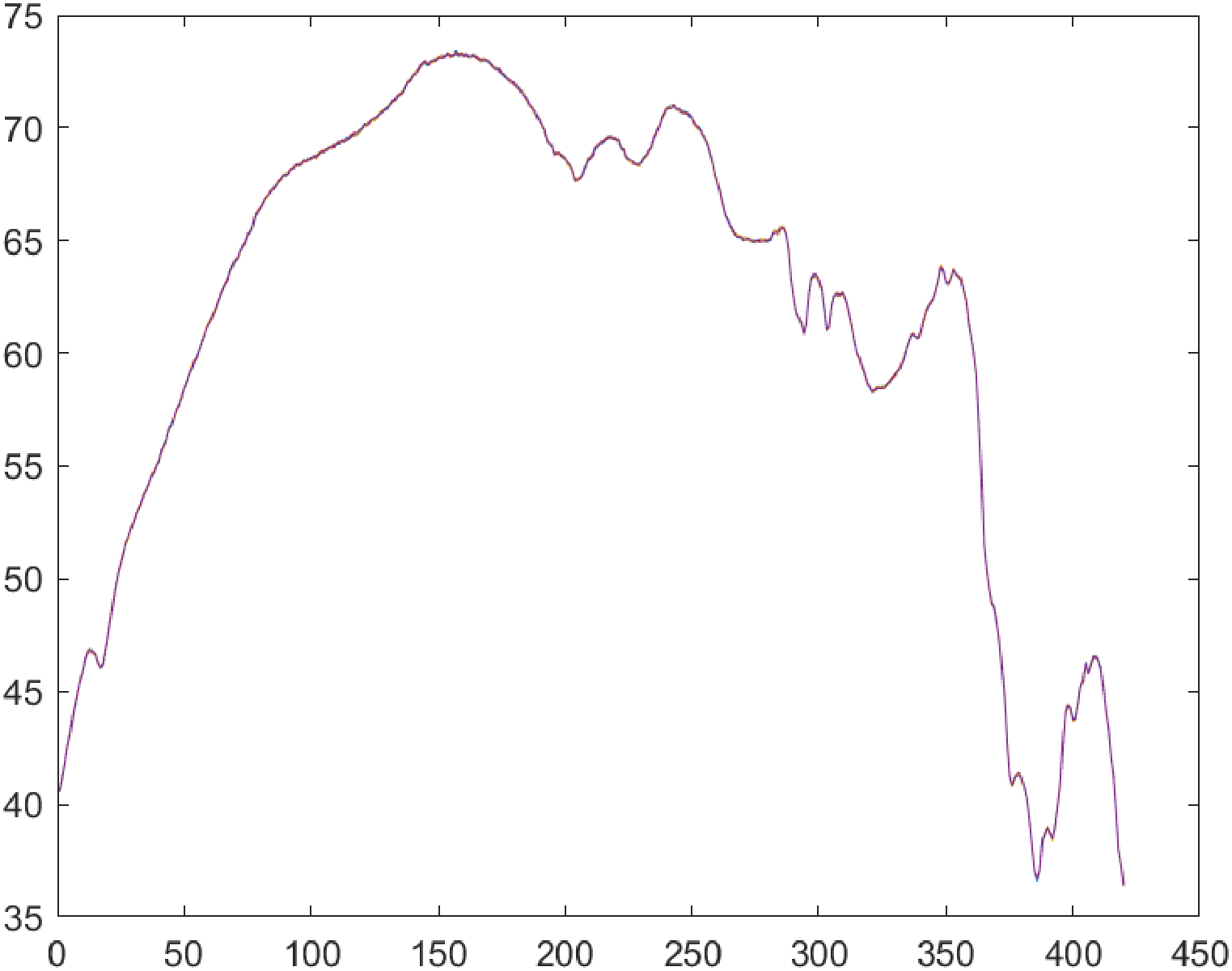}}
\end{center}
\caption{The Endmember Library and Spectral Variability Library obtained of the Synthetic data set through SVASU.}
\label{syncurve}
\end{figure}

\begin{figure}[t!]
\centering
\subfigure[]{\includegraphics[width=9cm]{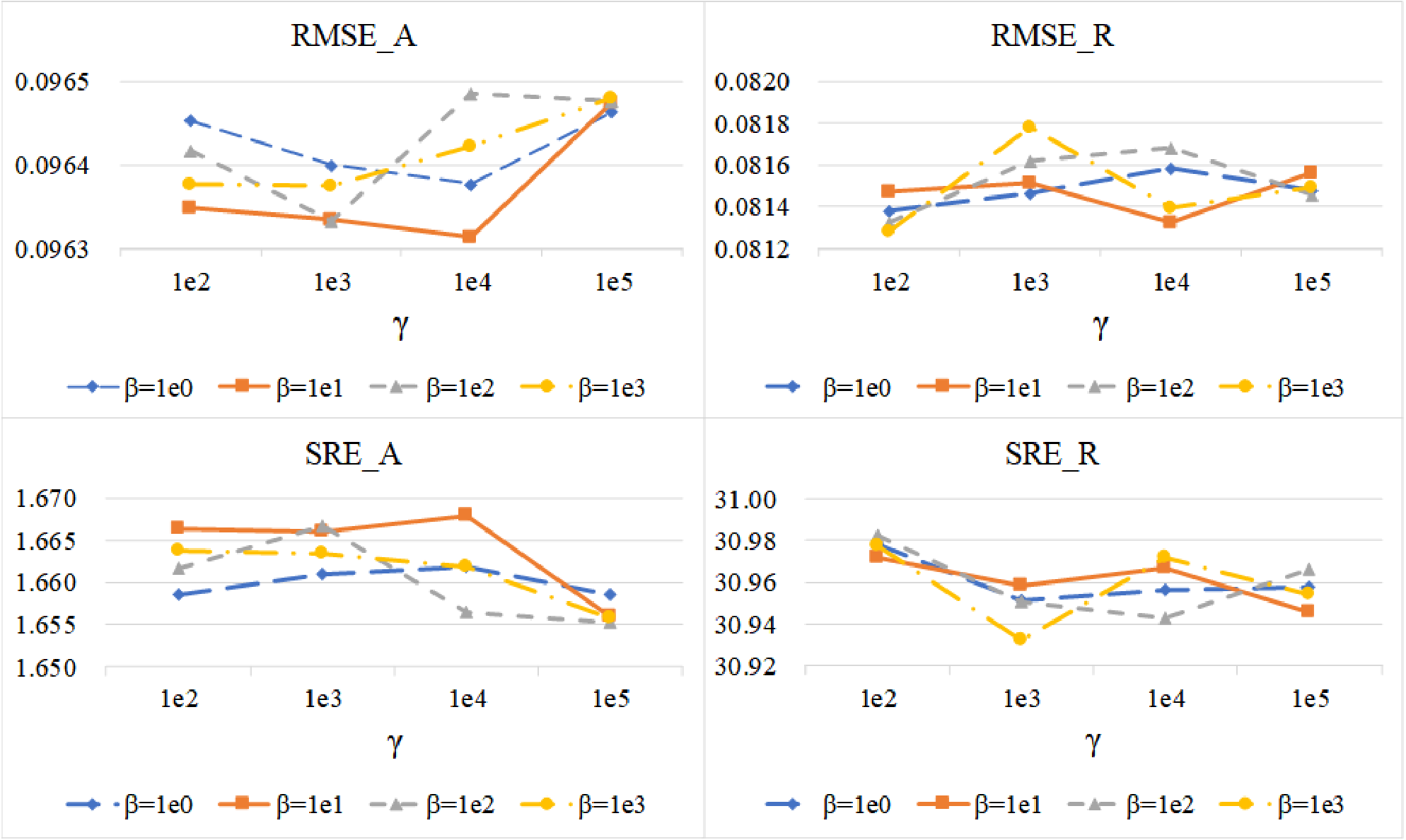}}
\subfigure[]{\includegraphics[width=9cm]{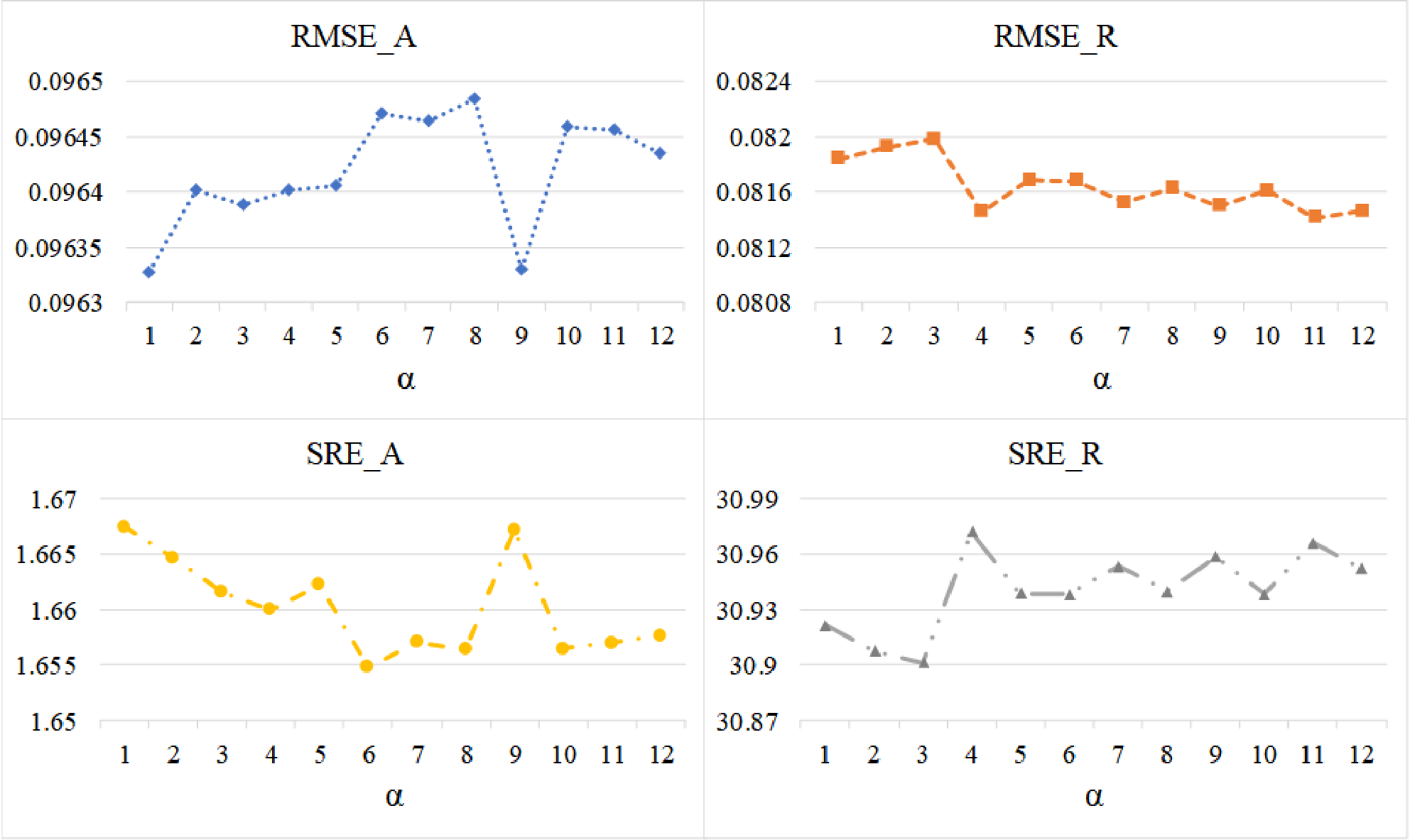}}
\caption{The performances of SVASU with respect to parameters (a) $\beta$, $\gamma$ and (b) $\alpha$ in terms of SRE and RMSE for both abundance and pixels on the synthetic data set.}
\label{synparam}
\end{figure}

\begin{table}[t!]
\centering
\label{synresult}
\caption{Average SRE value and RMSE value obtained by comparison algorithms on the Synthetic data set. Note that the best results are in bolded.}
\begin{tabular}{ccccccc}
\hline\hline
            & SUnSAL               & CLSUnSAL                      & JSpBLRU              & ADSpLRU                       & MUA$_{\rm{SLIC}} $            & \textbf{SVASU}  \\\hline\hline
SRE$_R$  & 8.0104    & 8.0026   & 7.1616          &8.0026 & 18.9487& \textbf{34.3370}\\
RMSE$_R$ & 1.2232    & 1.2241   & 1.3436          & 0.0608          & 0.5680 & \textbf{0.0573}\\
SRE$_A$  & 0.2409    & 0.2220   & 1.0789          & 0.1460          & 0.2334          & \textbf{1.7884} \\
RMSE$_A$ & 0.1136    & 0.1133   &0.0901 & 0.1129          & 0.1136          & \textbf{0.0732}\\\hline\hline
\end{tabular}
\end{table}

\subsection{Experiments Over Synthetic Data set}

In this section, the simulated data are generated in the way similar to \cite{Zhuang2019QMV} and \cite{Bioucas2012Overview} based on the LMM in (\ref{eqn_LMM}).
The synthetic data set has $n = 10000$ pixels. The endmember spectral signatures are using five spectra of minerals from the U.S. Geological Survey (USGS) containing as many as 420 bands covering from 400 to 2500 nm. Then, 30 dB Gaussian noise is added to each signature to simulate spectral variability.
Considering that each pixel is unlikely to have a large number of endmembers in real scenarios (typically, less than 5), the maximum number of active endmembers per pixel is set to 4 to ensure abundance sparsity. Furthermore, the abundance vectors corresponding to active endmembers are generated using a mixture of uniform Dirichlet distributions \cite{Jose2012Dirichlet}.
Finally, 40 dB Gaussian noise is added to the whole image.
In this experiment, the In-situ spectral library is simply composed of the selected five spectra from USGS and their corresponding spectra owning spectral variability. After the library segmentation by PCA, we acquire the (a) endmember library and (b) spectral variability library as shown in Fig. \ref{syncurve}.

First of all, the impact of parameters, including $\alpha$, $\beta$, and $\gamma$, are discussed through several sets of simulated experiments.
The curves of these parameters of our method are shown in Fig. \ref{synparam}(a) and (b). The evaluation metrics adopted here are the average SRE and RMSE calculated by (\ref{ReconEr1})-(\ref{SRE2}).
On one hand,
in Fig. \ref{synparam}(a), $\alpha$ is preset to 10 and $\beta$ and $\gamma$ vary in $[1e0,1e3]$ and $[1e2,1e5]$ exponentially.
The proposed SVASU algorithm achieves a desired result when $\beta =10$ and $\gamma = 1e4$, with lower RMSE value and higher SRE value both for abundance and pixels.
On the other hand, Fig. \ref{synparam}(b) displays the performances of SRE and RMSE under different values of parameter $\alpha$ tend to prominent when $\alpha$ reaches 9. Thus, the parameter $\alpha$, $\beta$, and $\gamma$ are respectively set to be 9, 10, and $1e4$ in the experiments on synthetic data set.

Table I shows the average SRE and RMSE value between the ground-truth pixels or abundance and corresponding matrixes acquired by different algorithms over synthetic data set. The performance of the proposed SVASU algorithm outperforms other methods on the estimation of either fractional abundance or pixel reconstruction.

\subsection{Experiments Over Jasper Data set}

\begin{figure}[t!]
\centering
\includegraphics[width=9cm]{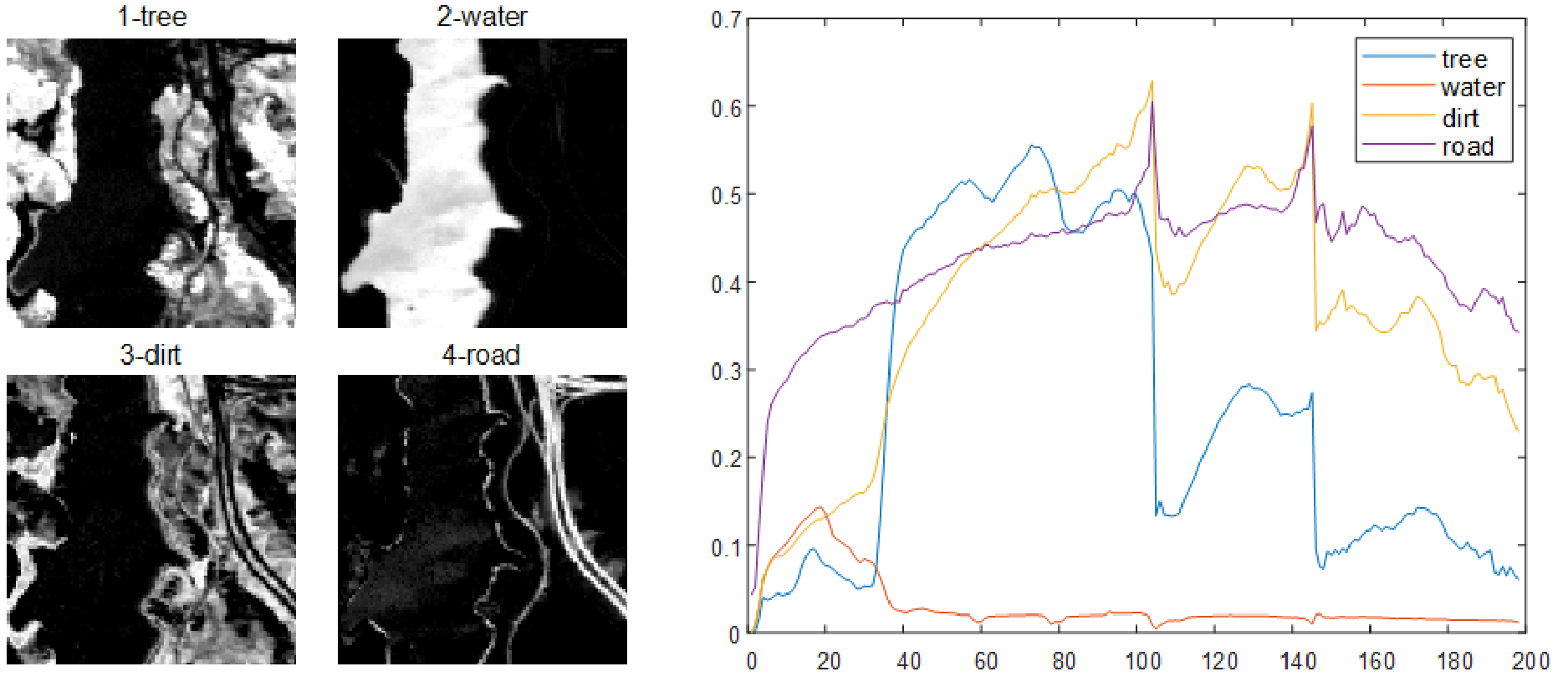}
\caption{The reference four endmembers and their corresponding abundance of the Jasper data set.}
\label{realimage}
\end{figure}

The real-world Jasper data set is collected by an airborne visible/infrared imaging spectrometer (AVIRIS) sensor\footnote{Data available online at https://rslab.ut.ac.ir/data.}, which consists of 224 spectral bands ranging from 380 to 2500 nm and the spectral resolution is up to 9.46 nm. The size of this scene is 512$\times$614. In the experiments, a subimage of 100$\times$100 pixels is used. Due to dense water vapor and atmospheric effects, there exists some noisy bands, including the bands 1$-$3, 108$-$112, 154$-$166, and 220$-$224. After removing these bands, a total of 198 reflectance bands are finally adopted. There are four endmembers in the selected region, including road, soil, water, and tree, and their corresponding ground-truth abundance are shown in Fig. \ref{realimage}.

\begin{figure}[t!]
\begin{center}
\subfigure[Endmember Library]{\includegraphics[width=4cm]{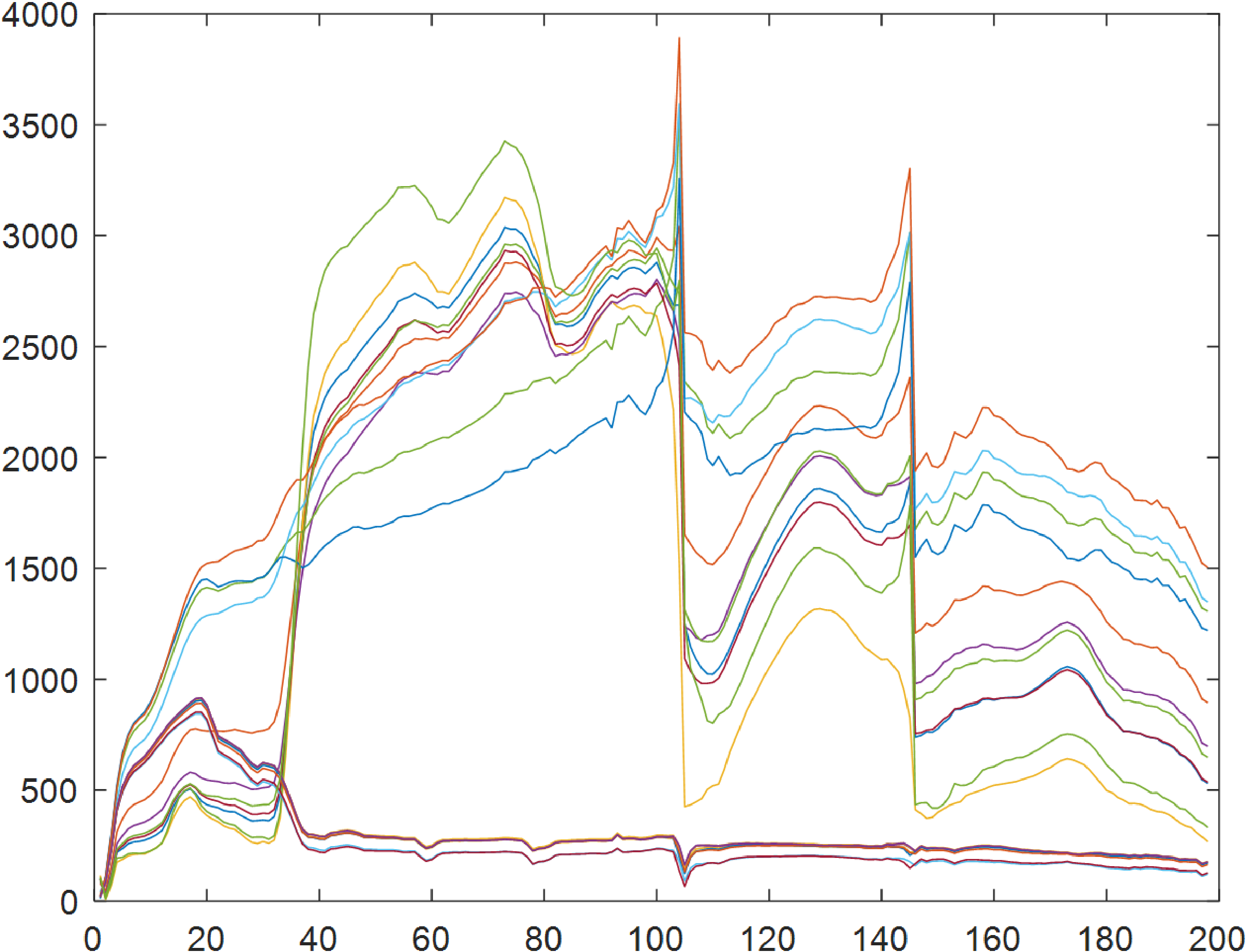}}
\subfigure[Spectral Variability Library]{\includegraphics[width=4cm]{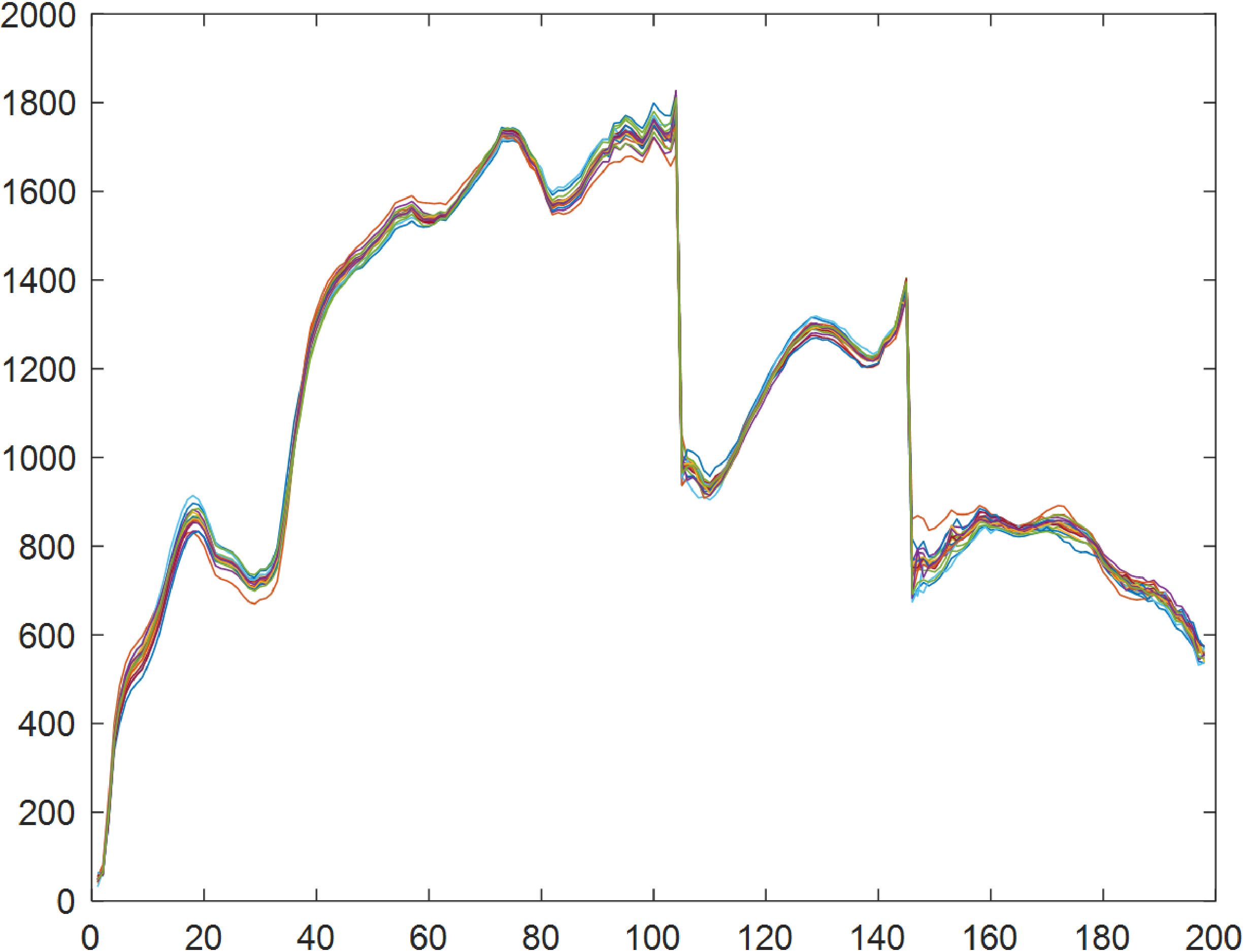}}
\end{center}
\caption{The Endmember Library and Spectral Variability Library obtained of the Jasper data set through SVASU.}
\label{realcurve}
\end{figure}


\begin{table}[t!]
\centering
\caption{Average SRE value and RMSE value obtained by comparison algorithms on the Jasper data set. Note that the best results are in bolded.}
\label{real}
\begin{tabular}{ccccccc}
\hline\hline
            & SUnSAL                     & CLSUnSAL                      & JSpBLRU              & ADSpLRU                       & MUA$_{\rm{SLIC}}$            & \textbf{SVASU}  \\\hline\hline
SRE$_R$  &15.8183                      & 16.9954                       & 18.7103              &17.2011             &18.6439         & \textbf{19.5641}         \\
RMSE$_R$ & 0.0986                        & 0.1083                        & 0.1024               & 0.1055                        &0.0789 & \textbf{0.0753}          \\
SRE$_A$  & 3.9046                        & 4.1053                        & 2.9957               & 2.6846                        & 6.7230          & \textbf{8.1455} \\
RMSE$_A$ & 0.1150                        & 0.0968                        & 0.1161               & 0.1224                        & 0.0796          & \textbf{0.0627} \\\hline\hline
\end{tabular}
\end{table}

\begin{figure}[htp]
\centering
\includegraphics[width=8cm]{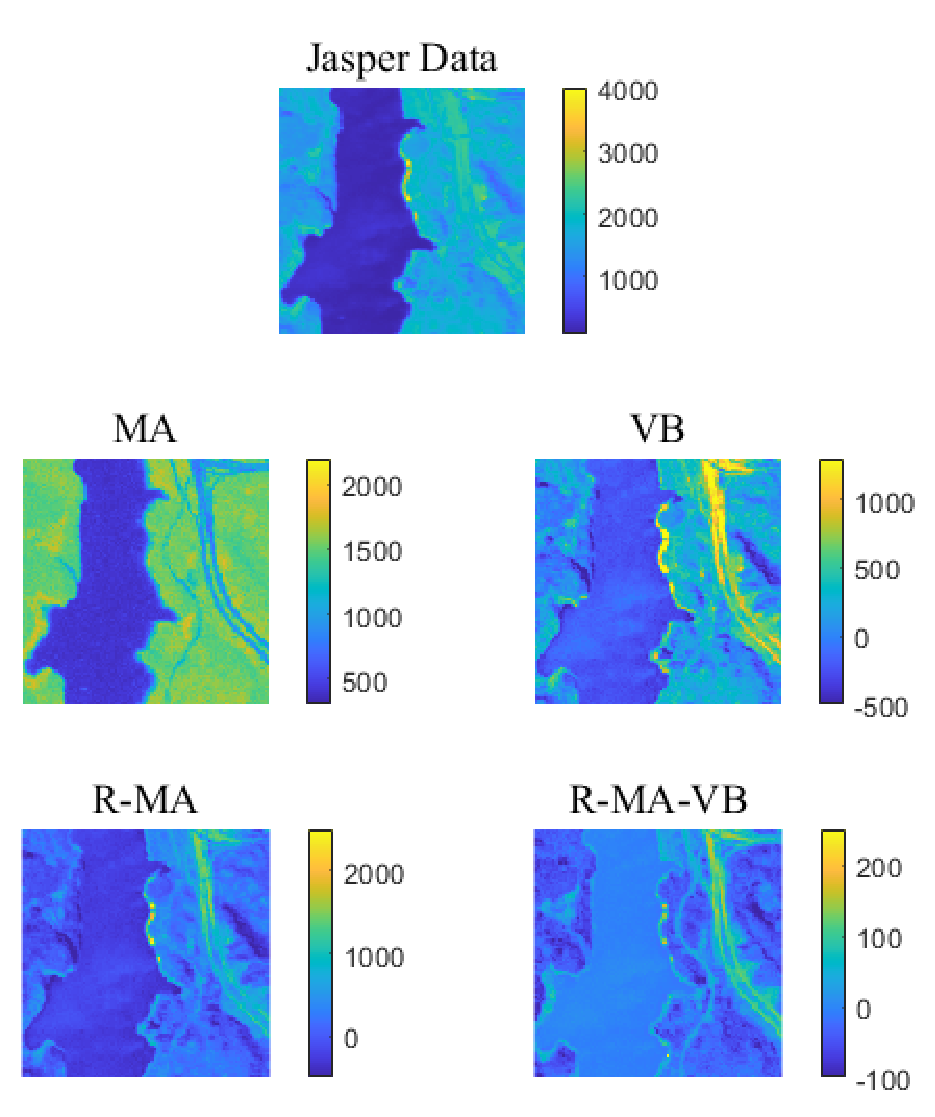}
\caption{Visual analysis of the proposed SVASU model conducted on Jasper data set.}
\label{visualresult1}
\end{figure}

As shown in Fig. 1, the In-situ spectral library obtained by SPEE is shown in the first column, and then after the segmentation step described in Section III. B, we get the endmember library and spectral variability library shown in Fig. \ref{realcurve}.
To analyze the impact of three regularization parameters, including $\alpha$, $\beta$, and $\gamma$, we first fix $\alpha$ at 10 to discuss the performance of SVASU under different values of $\beta$ and $\gamma$. As shown in Fig. X(a), with the increase in $\beta$ value, the performances of abundance estimation and pixel reconstruction demonstrate different trends of development.
Since the gap between the maximum and minimum values of evaluation matrixes for pixel reconstruction is small, the optimal parameter $\beta$ is set to $1e3$ and $\gamma$ is set to $1e6$ corresponds to good capability of abundance estimation.
After that, we further discuss the range of $\alpha$ as shown in Fig. X(b). Through observation, it can be found that when the value of $\alpha$ is lower than 3, the unmixing results of both abundance accuracy and data fidelity are superior to other cases. Thus, the optimal value of $\alpha$ is set to 1.

Table II shows the average SRE and RMSE values between the ground-truth pixels or abundance and corresponding matrixes acquired by different algorithms over synthetic data set.
The performance of the proposed SVASU algorithm outperforms other methods on the estimation of fractional abundance.
However, due to the second level decomposition by our extracted spectral variability, the performance of pixel reconstruction is not as well as other sparse unmixing methods as expected.

Furthermore, to carefully analyze the results of two-order reconstruction, Fig. \ref{visualresult1} depicts the visual results of the important items.
Each visual result is averaged over each band, including the image data $\mathbf{R}$, the first order reconstruction item $\mathbf{MA}$, the second order reconstruction item $\mathbf{VB}$, the first order reconstruction error item $\mathbf{R-MA}$, and the second order reconstruction error item $\mathbf{R-MA-VB}$.
It can be observed that on the Jasper data set, the amount of information occupied by $\mathbf{MA}$ is probably twice as much information as that represents by $\mathbf{VB}$.
Moreover, through the second order reconstruction by spectral variability library, the image reconstruction error reduced to a tenth on Jasper data set, which completely proves the validity and rationality of the proposed SVASU method.

\subsection{Experiments Over Cuprite Data set}

\begin{figure}[htp]
\centering
\includegraphics[width=8cm]{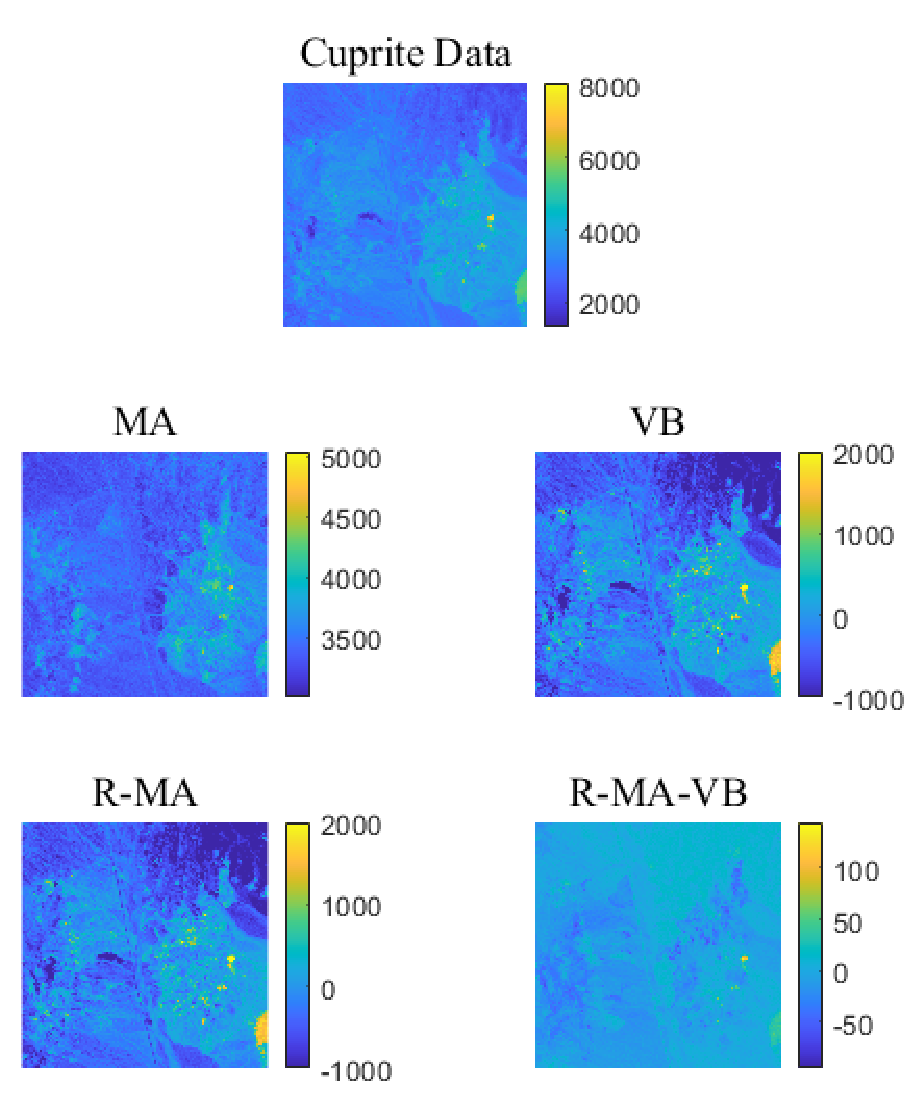}
\caption{Visual analysis of the proposed SVASU model conducted on Cuprite data set.}
\label{visualresult2}
\end{figure}

The hyperspectral data set named Cuprite data set, acquired by AVIRIS\footnote{http://aviris.jpl.nasa.gov/html/aviris.freedata.html.}, is also adopted in real experiments, which contains 224 bands ranging from 370 to 2510 nm with a ground instantaneous field of view of 20 m.
The cropped image corresponds to a $350\times350$ pixel subset of the sector labeled as $f970619t01p02_r02_sc03.a.rfl$ in the online data.
After removing noisy bands and water-absorption bands (including bands $1-4$, $105-115$, $150-170$, and $223-224$), a total of 186 reflectance bands are finally adopted. The optimal parameters of proposed SVASU method are $\alpha=3$, $\beta=1e4$, and $\gamma=1e6$.

Given that there is no available ground-truth in terms of abundances and endmembers of Cuprite data set, only visual results of two-order reconstruction in SVASU are shown as Fig. \ref{visualresult2}.
Similarly, each visual result is averaged over each band, including the two-order reconstruction item $\mathbf{MA}$, $\mathbf{VB}$, and the two-order reconstruction error item $\mathbf{R-MA}$, $\mathbf{R-MA-VB}$.
It can be observed that the amount of information occupied by $\mathbf{MA}$ is more than twice as much information as that by $\mathbf{VB}$.
This also confirms that in the processing of $PCA$, the main components are constructed by endmember library while the other components are constructed as the spectral variability library.
Moreover, through the second order reconstruction by spectral variability library, the image reconstruction error reduced to less than a tenth on Cuprite data set.
Therefore, the effectiveness of the proposed SVASU method are confirmed on the real-world data set.

\section{Conclusion}
In this paper, a spectral variability augmented sparse unmixing model (SVASU) is proposed, which provides a new way to construct spectral variability explicitly from the spectral library for sparse unmixing. It is noted that a in-situ spectral library acquired under spatial purity is separated into the endmember library and the variability library by PCA. Afterwards, the proposed SVASU adopts a two-order decomposition structure to perform SU, in which the first order is to decompose the hyperspectral image data into the product of endmember library and abundance fractions roughly, the second order is to further represent the reconstruction error of pixels from first order as the influence of spectral variability library.
Experimental results over both synthetic and real-world datasets demonstrate that the proposed SVASU model can certainly improve the performance of spectral linearly unmixing and through the introducing of second order reconstructed loss, the ability of the proposed framework to model the spectral variability is improved. Meanwhile, the use of those smoothness terms guarantees a moderate variation of the spectral variability.

\appendices
\section{Proof of Convergence of SVASU}

The problem in (\ref{eqn_SVASU1}) contains two unknown matrices, and we design different rules to alternately update one by fixing the other.
Following \cite{Lee2001NMF,Huang2019Auxi}, $\emph {auxiliary function approach}$ is used to prove the convergence of the proposed update rules.
Function $Q(\mathbf{X},\mathbf{\tilde X})$ is an auxiliary function for $L(\mathbf{X})$ if the conditions
\begin{equation}
\begin{aligned}
\label{prove1}
Q(\mathbf{X},\mathbf{\tilde X})\geq L(\mathbf{X}),\ Q(\mathbf{X},\mathbf{X})= L(\mathbf{X})
\end{aligned}
\end{equation}
are satisfied for any $\mathbf{X},\mathbf{\tilde X}$. Then, if
\begin{equation}
\begin{aligned}
\label{prove2}
\mathbf{X}^{(t+1)}= \mathop {arg\min\limits_\mathbf{X}} Q(\mathbf{X},\mathbf{X}^{(t)}),
\end{aligned}
\end{equation}
it has been proven that the following inequalities held:
\begin{equation}
\begin{aligned}
\label{prove3}
L(\mathbf{X}^{(t+1)})\leq Q(\mathbf{X}^{(t+1)},\mathbf{X}^{(t)})\leq Q(\mathbf{X}^{(t)},\mathbf{X}^{(t)})=L(\mathbf{X}^{(t)}).
\end{aligned}
\end{equation}
Thus, the function $L(\mathbf{X})$ is monotonically decreasing.

\subsubsection{Fixing $\mathbf{B}$, minimizing $\mathbf{A}$}
First, when $\mathbf{B}$ is fixed, the minimization problem in (\ref{eqn_SVASU1}) can be written as

\begin{equation}
\begin{aligned}
\label{fixB}
\mathop {\min }\limits_\mathbf{A} {\left\| {\mathbf{R - MA}} \right\|_F^2} + \alpha {\left\| \mathbf{T - MA} \right\|_F^2} + \beta Tr(\mathbf{A}^T\mathbf{DA})\\
{s.t.\mathbf{A} > 0,}
\end{aligned}
\end{equation}
in which $\mathbf{T} = \mathbf{R} - \mathbf{VB}$.
For the first two terms in (\ref{fixB}), the following inequality holds \cite{Wang2011Community}.
\begin{equation}
\begin{aligned}
{L_1}(\mathbf{A}) &= \left\| {\mathbf{R} - \mathbf{M}\mathbf{A}} \right\|_F^2\\
 &= Tr({\mathbf{R}^T}\mathbf{R}) - 2Tr({\mathbf{R}^T}\mathbf{M}\mathbf{A}) + Tr({\mathbf{A}^T}{\mathbf{M}^T}\mathbf{M}\mathbf{A})\\
 &\leq Tr({\mathbf{R}^T}\mathbf{R}) - 2Tr({\mathbf{R}^T}\mathbf{M}\mathbf{\tilde A})-2Tr({\mathbf{R}^T}\mathbf{MZ})\\
 &+ \frac{1}{2}Tr({\mathbf{P}^T}{\mathbf{M}^T}\mathbf{M}\mathbf{\tilde A}) + \frac{1}{2}Tr({\mathbf{\tilde A}^T}{\mathbf{M}^T}\mathbf{M}\mathbf{\tilde A})\\
 &={Q_1}(\mathbf{A},\mathbf{\tilde A})
\end{aligned}
\end{equation}
\begin{equation}
\begin{aligned}
{L_2}(\mathbf{A}) &= \left\| {\mathbf{T} - \mathbf{M}\mathbf{A}} \right\|_F^2\\
 &= Tr({\mathbf{T}^T}\mathbf{T}) - 2Tr({\mathbf{T}^T}\mathbf{M}\mathbf{A}) + Tr({\mathbf{A}^T}{\mathbf{M}^T}\mathbf{M}\mathbf{A})\\
 &\leq Tr({\mathbf{T}^T}\mathbf{T}) - 2Tr({\mathbf{T}^T}\mathbf{M}\mathbf{\tilde A})-2Tr({\mathbf{T}^T}\mathbf{MZ})\\
 &+ \frac{1}{2}Tr({\mathbf{P}^T}{\mathbf{M}^T}\mathbf{M}\mathbf{\tilde A})+ \frac{1}{2}Tr({\mathbf{\tilde A}^T}{\mathbf{M}^T}\mathbf{M}\mathbf{\tilde A}) \\
 &={Q_2}(\mathbf{A},\mathbf{\tilde A})
\end{aligned}
\end{equation}
where $\mathbf{P}_{ik}=(\mathbf{A}_{ik})^4/(\mathbf{\tilde A}_{ik})^3$, and $\mathbf{Z}_{ik}=\mathbf{\tilde A}_{ik}\ln(\mathbf{A}_{ik}/\mathbf{\tilde A}_{ik})$.
It is obvious that ${Q_1}(\mathbf{A},\mathbf{A})={L_1}(\mathbf{A}), {Q_2}(\mathbf{A}, \mathbf{A})={L_2}(\mathbf{A})$.
Taking $Q_1(\mathbf{A},\mathbf{\tilde A})$ as an example, $Q_1(\mathbf{A},\mathbf{\tilde A})\geq L_1(\mathbf{A})$ can be proved in detail as follows,
\begin{equation}
\begin{aligned}
&Q_1(\mathbf{A},\mathbf{\tilde A})-L_1(\mathbf{A})\\
&= 2Tr({\mathbf{R}^T}\mathbf{M}\mathbf{A})-2Tr({\mathbf{R}^T}\mathbf{M}\mathbf{\tilde A})-2Tr({\mathbf{R}^T}\mathbf{MZ})\\
&+ \frac{1}{2}Tr({\mathbf{P}^T}{\mathbf{M}^T}\mathbf{M}\mathbf{\tilde A}) + \frac{1}{2}Tr({\mathbf{\tilde A}^T}{\mathbf{M}^T}\mathbf{M}\mathbf{\tilde A})\\
&- Tr({\mathbf{A}^T}{\mathbf{M}^T}\mathbf{M}\mathbf{A})\\
&=\sum\limits_{ik} 2{\mathbf{R}^T_{ik}}{\mathbf{M}_{ik}}({\mathbf{A}_{ik}}-{\mathbf{\tilde A}}_{ik}-{\mathbf{\tilde A}}_{ik}\ln \frac{\mathbf{A}_{ik}}{\mathbf{\tilde A}_{ik}})\\
&+\sum\limits_{ik}\frac{1}{2}({\mathbf{M}_{ik}})^2(\frac{({\mathbf{A}_{ik}})^2}{{{\mathbf{\tilde A}}_{ik}}}-{\mathbf{\tilde A}_{ik}})^2 \geq 0.
\end{aligned}
\end{equation}
Thus, $Q_1(\mathbf{A},\mathbf{\tilde A})$ and $Q_2(\mathbf{A},\mathbf{\tilde A})$ can be selected as the auxiliary functions of $L_1(\mathbf{A})$ and $L_2(\mathbf{A})$ correspondingly.

Then, for the last term $L_3(\mathbf{A})=Tr({\mathbf{A}^T}\mathbf{D}\mathbf{A})$, the following auxiliary function $Q_3(\mathbf{A},\mathbf{\tilde A})$ is constructed as,
\begin{equation}
\begin{aligned}
Q_3(\mathbf{A},\mathbf{\tilde A})=\frac{1}{2}Tr({\mathbf{P}^T}\mathbf{D}\mathbf{\tilde A}) + \frac{1}{2}Tr({{\mathbf{\tilde A}}^T}\mathbf{D}\mathbf{\tilde A}),
\end{aligned}
\end{equation}
It is straightforward to verify that $Q_3(\mathbf{A},\mathbf{A})=L_3(\mathbf{A})$. To show that $Q_3(\mathbf{A},\mathbf{\tilde A})\geq L_3(\mathbf{A})$, we have the following calculation,
\begin{equation}
\begin{aligned}
&Q_3(\mathbf{A},\mathbf{\tilde A})-L_3(\mathbf{A})\\
&=\sum\limits_{ik} {\frac{{{\mathbf{D}_{ii}}{{\mathbf{\tilde A}}_{ik}}{{({\mathbf{A}_{ik}})}^4}}}{{{{2 (\mathbf{\tilde A}}_{ik})^3}}}}  + \sum\limits_{ik} { \frac{{{\mathbf{D}_{ii}}}} {2}({{\mathbf{\tilde A}}_{ik}})^2  - \sum\limits_{ik} {{\mathbf{D}_{ii}}({\mathbf{A}_{ik}})}^2} \\
&=\sum\limits_{ik} \frac{{{\mathbf{D}_{ii}}}} {2}(\frac{({\mathbf{A}_{ik}})^4}{({{\mathbf{\tilde A}}_{ik}})^2}-2 ({\mathbf{A}_{ik}})^2 + ({{\mathbf{\tilde A}}_{ik}})^2) \\
&=\sum\limits_{ik} \frac{{\mathbf{D}_{ii}}} {2}(\frac{({\mathbf{A}_{ik}})^2}{\mathbf{\tilde A}_{ik}}-{\mathbf{\tilde A}_{ik}})^2  \geq 0.
\end{aligned}
\end{equation}
Thus, $Q_3(\mathbf{A},\mathbf{\tilde A})$ can be an auxiliary function of $L_3(\mathbf{A})$.

Finally, $Q(\mathbf{A},\mathbf{\tilde A})=Q_1(\mathbf{A},\mathbf{\tilde A})+ \alpha Q_2(\mathbf{A},\mathbf{\tilde A})+\beta Q_3(\mathbf{A},\mathbf{\tilde A})$ is an auxiliary function of $L(\mathbf{A})=L_1(\mathbf{A})+\alpha L_2(\mathbf{A})+ \beta L_3(\mathbf{A})$ in (\ref{eqn_SVASU1}).
Currently, according to Eq. (\ref{prove2}), we find the minimum of $\min Q(\mathbf{A},\mathbf{\tilde A})$ by fixing $\mathbf{\tilde A}$, which is given by
\begin{equation}
\begin{aligned}
0 &= \frac{{\partial Q(\mathbf{A},\mathbf{\tilde A})}}{{\partial {\mathbf{A}_{ik}}}} \\
&= \frac{{\partial {Q_1}(\mathbf{A},\mathbf{\tilde A})}}{{\partial {\mathbf{A}_{ik}}}} + \alpha \frac{{\partial {Q_2}(\mathbf{A},\mathbf{\tilde A})}}{{\partial {\mathbf{A}_{ik}}}} + \beta \frac{{\partial {Q_3}(\mathbf{A},\mathbf{\tilde A})}}{{\partial {\mathbf{A}_{ik}}}}\\
 &= 2(1+\alpha)\frac{({\mathbf{A}_{ik}})^3}{({{\mathbf{\tilde A}}_{ik}})^3}({\mathbf{M}^T}\mathbf{M\tilde A})_{ik} - 2\frac{{{{\mathbf{\tilde A}}_{ik}}}}{{{\mathbf{A}_{ik}}}}{({\mathbf{M}^T}\mathbf{R})_{ik}} \\
 &- 2\alpha \frac{{{{\mathbf{\tilde A}}_{ik}}}}{{{\mathbf{A}_{ik}}}}({\mathbf{M}^T}\mathbf{T})_{ik} + 2\beta \frac{{{{({\mathbf{A}_{ik}})}^3}}}{({{\mathbf{\tilde A}}_{ik}})^3}{(\mathbf{D\tilde A})_{ik}}.
\end{aligned}
\end{equation}
Then, we can get
\begin{equation}
\begin{aligned}
\label{ruleA}
&\mathbf{A}_{ik} =\mathbf{\tilde A}_{ik} (\frac{{({\mathbf{M}^T}\mathbf{R} + \alpha \mathbf{M}^T\mathbf{R}- \alpha \mathbf{M}^T\mathbf{VB} )_{ik}}}{{((1 + \alpha ){\mathbf{M}^T}\mathbf{M}\mathbf{\tilde A} + \beta \mathbf{D} \mathbf{\tilde A})_{ik}}})^ \frac{1}{4}.
\end{aligned}
\end{equation}
\subsubsection{Fixing $\mathbf{A}$, minimizing $\mathbf{B}$}
When $\mathbf{A}$ is fixed, the minimization problem in (\ref{eqn_SVASU1}) can be written as
\begin{equation}
\begin{aligned}
\label{fixA}
\mathop {\min }\limits_\mathbf{B} \alpha{\left\| {\mathbf{T' - VB}} \right\|_F^2} + \gamma {\left\| {\mathbf{B}} \right\|_F^2}\ \ \ {s.t.\mathbf{B} > 0,}
\end{aligned}
\end{equation}
in which $\mathbf{T'} = \mathbf{R} - \mathbf{MA}$.

Similarly, the following two inequalities hold:
\begin{equation}
\begin{aligned}
{L_1}(\mathbf{B}) &= \left\| {\mathbf{T'} - \mathbf{V}\mathbf{B}} \right\|_F^2\\
 &= Tr({\mathbf{T'}^T}\mathbf{T'}) - 2Tr({\mathbf{T'}^T}\mathbf{V}\mathbf{B}) + Tr({\mathbf{B}^T}{\mathbf{V}^T}\mathbf{V}\mathbf{B})\\
 &\leq Tr({\mathbf{T'}^T}\mathbf{T'}) - 2Tr({\mathbf{T'}^T}\mathbf{V}\mathbf{\tilde B})-2Tr({\mathbf{T'}^T}\mathbf{VZ'})\\
 &+\frac{1}{2} Tr({\mathbf{P'}^T}{\mathbf{V}^T}\mathbf{V}\mathbf{\tilde B})+\frac{1}{2} Tr({\mathbf{\tilde B}^T}{\mathbf{V}^T}\mathbf{V}\mathbf{\tilde B}) \\
 &={Q_1}(\mathbf{B},\mathbf{\tilde B})
\end{aligned}
\end{equation}
\begin{equation}
\begin{aligned}
L_2(\mathbf{B})&=\left\| {\mathbf{B}} \right\|_F^2=Tr({\mathbf{B}^T}\mathbf{B})\\
&\leq \frac{1}{2}Tr(\mathbf{P'}^T \mathbf{\tilde B})+ \frac{1}{2}Tr(\mathbf{\tilde B}^T \mathbf{\tilde B}) = {Q_2}(\mathbf{B},\mathbf{\tilde B})
\end{aligned}
\end{equation}
in which $\mathbf{P'}_{ij}=(\mathbf{B}_{ij})^4/(\mathbf{\tilde B}_{ij})^3$, and $\mathbf{Z'}_{ij}=\mathbf{\tilde B}_{ij}\ln(\mathbf{B}_{ij}/\mathbf{\tilde B}_{ij})$.

Thus, $Q(\mathbf{B},\mathbf{\tilde B})=\alpha Q_1(\mathbf{B},\mathbf{\tilde B})+  \gamma Q_2(\mathbf{B},\mathbf{\tilde B})$ is an auxiliary function of $L(\mathbf{B})= \alpha L_1(\mathbf{B})+  \gamma L_2(\mathbf{B})$ in Eq. (\ref{fixA}).
Currently, according to Eq. (\ref{prove2}), we find the minimum of $\min Q(\mathbf{B},\mathbf{\tilde B})$ by fixing $\mathbf{\tilde B}$, which is given by
\begin{equation}
\begin{aligned}
0 &= \frac{{\partial Q(\mathbf{B},\mathbf{\tilde B})}}{{\partial {\mathbf{B}_{ij}}}} \\
&= \alpha \frac{{\partial {Q_1}(\mathbf{B},\mathbf{\tilde B})}}{{\partial {\mathbf{B}_{ij}}}} + \gamma \frac{{\partial {Q_2}(\mathbf{B},\mathbf{\tilde B})}}{{\partial {\mathbf{B}_{ij}}}}\\
 &=  2\alpha \frac{({\mathbf{B}_{ij}})^3}{(\mathbf{\tilde B}_{ij})^3} ({\mathbf{V}^T}\mathbf{V}\mathbf{\tilde B})_{ij}-2\alpha  \frac{{\mathbf{\tilde B}_{ij}}}{\mathbf{B}_{ij}}({\mathbf{V}^T}\mathbf{T'})_{ij} \\
 &+ 2\gamma \frac{({\mathbf{B}_{ij}})^3}{(\mathbf{\tilde B}_{ij})^2}
\end{aligned}
\end{equation}
Solving for $\mathbf{B}_{ij}$, the minimum is
\begin{equation}
\begin{aligned}
\label{ruleB}
{\mathbf{B}_{ij}} = \mathbf{\tilde B}_{ij}  (\frac{{(\alpha {\mathbf{V}^T}\mathbf{R}- \alpha {\mathbf{V}^T}\mathbf{MA})_{ij}}}{{( \alpha {\mathbf{V}^T}\mathbf{V}\mathbf{\tilde B} + \gamma \mathbf{\tilde B})_{ij}}})^ \frac{1}{4}.
\end{aligned}
\end{equation}

Thus, updating $\mathbf{A}$ and $\mathbf{B}$ alternatively by using the rules in (\ref{ruleA}) and (\ref{ruleB}) can guarantee that the objective function in (\ref{eqn_SVASU1}) is non-increasing.


\bibliographystyle{IEEEtran}
\bibliography{SVASU}

\end{document}